\documentclass{article}
\usepackage{amsmath}
\usepackage{mathtools}
\usepackage{graphicx}
\usepackage{subfigure}
\usepackage{titling}
\usepackage{authblk}
\usepackage{hyperref}
\usepackage{etoolbox}
\usepackage{lmodern}
\usepackage{tikz}
\usepackage[T1]{fontenc}
\usepackage{multicol}
\usepackage{CJK}
\usepackage{algorithm}
\usepackage{algorithmic}
\usepackage{braket}
\usepackage[utf8]{inputenc}
\usepackage{fancyhdr}
\usepackage[font=small,labelfont=bf]{caption} 
\usepackage{geometry}
\newcommand\z[1]{\sigma^z_{#1}}
\newcommand\Z[1]{\sigma^{z'}_{#1}}

\begin{document}
\title{Quantum Machine Learning for Electronic Structure Calculations}
\author[1]{Rongxin Xia}
\author[1,2,3]{Sabre Kais \thanks{kais@purdue.edu}}
\affil[1]{Department of Physics and Astronomy, Purdue University, West Lafayette, IN, 47907 USA}
\affil[2]{Department of Chemistry and Birck Nanotechnology Center, Purdue University,
West Lafayette, IN 47907 USA}
\affil[3]{Santa Fe Institute, 1399 Hyde Park Rd, Santa Fe, NM 87501}
\date{}

\setlength{\parindent}{0pt}
\setlength{\parskip}{0.5\baselineskip}
\maketitle
\vspace{-8ex}

\begin{abstract}
\bf{Considering recent advancements and successes in the development of efficient quantum algorithms for electronic structure calculations --- alongside impressive results using machine learning techniques for computation --- hybridizing quantum computing with machine learning for the intent of performing electronic structure calculations is a natural progression. Here we report a hybrid quantum algorithm employing a restricted Boltzmann machine to obtain accurate molecular potential energy surfaces. By exploiting a quantum algorithm to help optimize the underlying objective function, we obtained an efficient procedure for the calculation of the electronic ground state energy for a small molecule system. Our approach achieves high accuracy for the ground state energy for H$_2$, LiH, H$_2$O at a specific location on its potential energy surface with a finite basis set.  With the future availability of larger-scale quantum computers, quantum machine learning techniques are set to become powerful tools to obtain accurate values for electronic structures.} 
\end{abstract}

\section*{Introduction}
Machine learning techniques are demonstrably powerful tools displaying remarkable success in compressing high dimensional data \cite{hinton2006reducing, lecun2015deep}. These methods have been applied to a variety of fields in both science and engineering, from computing excitonic dynamics  \cite{hase2016machine}, energy transfer in light-harvesting systems  \cite{hase2017machine}, molecular electronic properties \cite{montavon2013machine}, surface reaction network  \cite{ulissi2017address}, learning density functional models \cite{brockherde2017bypassing} to classify phases of matter, and the simulation of classical and complex quantum systems \cite{wang2016discovering,carrasquilla2017machine,broecker2017machine,ch2017machine,van2017learning,arsenault2014machine,kusne2014fly}. Modern machine learning techniques have been used in the state space of complex condensed-matter systems for their abilities to analyze and interpret exponentially large data sets \cite{carrasquilla2017machine} and to speed-up searches for novel energy generation/storage materials \cite{de2017use, wei2016neural}. 

Quantum machine learning \cite{biamonte2017quantum} - hybridization of classical machine learning techniques with quantum computation -- is emerging as a powerful approach allowing quantum speed-ups and improving classical machine learning algorithms 
 \cite{lloyd2013quantum, rebentrost2014quantum,neven2008image,neven2008training,neven2009training}. Recently, Wiebe \emph{et. al.} \cite{wiebe2016quantum} have shown that quantum computing is capable of reducing the time required to train a restricted Boltzmann machine (RBM), while also providing a richer framework for deep learning than its classical analogue. The standard RBM models the probability of a given configuration of visible and hidden units by the Gibbs distribution with interactions restricted between different layers. Here, we focus on an RBM where the visible and hidden units assume $\{+1,-1\}$ forms \cite{carleo2017solving,torlai2018neural}. 
 
 Accurate electronic structure calculations for large systems continue to be a challenging problem in the field of chemistry and material science. Toward this goal --- in addition to the impressive progress in developing classical algorithms based on \emph{ab initio} and density functional methods --- quantum computing based simulation have been explored \cite{kais2014introduction,daskin2018direct,aspuru2005simulated,o2016scalable,kassal2011simulating,babbush2014adiabatic}. Recently, Kivlichan \emph{et. al.} \cite{kivlichan2018quantum} show that using a particular arrangement of gates (a fermionic swap network) it is possible to simulate electronic structure Hamiltonian with linear depth and connectivity. These results present significant improvement on the cost of quantum simulation for both variational and phase estimation based quantum chemistry simulation methods. 

Recently, Troyer and coworkers proposed using a restricted Boltzmann machine to solve quantum many-body problems, for both stationary states and time evolution of the quantum Ising and Heisenberg models \cite{carleo2017solving}. However, this simple approach has to be modified for cases where the wave function's phase is required for accurate 
calculations \cite{torlai2018neural}.

Herein, we propose a three-layered RBM structure that includes the visible and hidden layers, plus 
a new layer correction for the signs of coefficients for basis functions of the wave function.
We will show that this model has the potential to solve complex quantum many-body problems and to obtain very accurate results for simple molecules as compared with the results calculated by a finite minimal basis set, STO-3G. We also employed a quantum algorithm to help the optimization of training procedure.

\section*{Results}

{\bf Three-layers restricted Boltzmann machine.} We will begin by briefly outlining the original RBM structure as described by  \cite{carleo2017solving}. For a given Hamiltonian, $H$, and a trial state, $|\phi\rangle = \sum_{x}\phi(x)|x\rangle$, the expectation value can be written as\cite{carleo2017solving}:
\begin{equation}
\langle H\rangle=\frac{\langle\phi|H|\phi\rangle}{\langle\phi|\phi\rangle}=\frac{\sum_{x,x'}\langle\phi|x\rangle \langle x|H|x'\rangle\langle x'|\phi\rangle}{\sum_{x}\langle\phi|x\rangle\langle x|\phi\rangle}=\frac{\sum_{x,x'}\overline{\phi(x)}\langle x|H|x'\rangle\phi(x')}{\sum_{x}|\phi(x)|^2}
\end{equation}
where $\phi(x) = \langle x|\phi\rangle$ will be used throughout this letter to express the overlap of the complete wave function with the basis function $|x\rangle$, $\overline{\phi(x)}$ is the complex conjugate of $\phi(x)$.

We can map the above to a RBM model with visible layer units $\z{1},\ \z{2}...\ \z{n}$ and hidden layer units $h_1,\ h_2...\ h_m $ with $\z{i}$, $h_j\in \{-1,1\}$. We use visible units $\z{i}$ to represent the spin state of a qubit $i$ -- up or down. The total spin state of $n$ qubits is represented by the basis $|x\rangle=|\z{1}\z{2}...\z{n}\rangle$. $\phi(x) = \sqrt{P(x)}$ where $P(x)$ is the probability for $x$ from the distribution determined by the RBM. The probability of a specific set $x=\{\z{1},\z{2}...\z{n}\}$ is:

\begin{equation}
P(x) = \frac{\sum_{\{h\}}e^{({\sum_{i}a_i\z{i}+\sum_{j}b_jh_j+\sum_{i,j}w_{ij}\z{i}h_j})}}{\sum_{x'}\sum_{\{h\}}e^{({\sum_ia_i\Z{i}+\sum_{j}b_jh_j+\sum_{i,j}w_{ij}\Z{i}h_j})}} 
\end{equation}

Within the above $a_i$ and $b_j$ are trainable weights for units $\z{i}$ and $h_j$. $w_{ij}$ are trainable weights describing the connections between $\z{i}$ and $h_j$ (see Figure 1.)

By setting $\langle H\rangle$ as the objective function of this RBM, we can use the standard gradient decent method to update parameters, effectively minimizing $\langle H\rangle$ to obtain the ground state energy.


However, previous prescriptions considering the use of RBMs for electronic structure problems have found difficulty as $\phi(x_i)$ can only be non-negative values. 
We have thus appended an additional layer to the neural network architecture to compensate for the lack of sign features specific to electronic structure problems.

We propose an RBM with three layers. The first layer, $\sigma^z$, describes the parameters building the wave function. The $h$'s within the second layer are parameters for the coefficients for the wave functions and the third layer $s$, represents the signs associated $|x\rangle$:

\begin{equation}
s(x) = s(\z{1}, \z{2}...\z{n}) = tanh(\sum_{i}d_i\z{i}+c)
\end{equation}

\begin{figure}[H]
\centering
\includegraphics[width=6in]{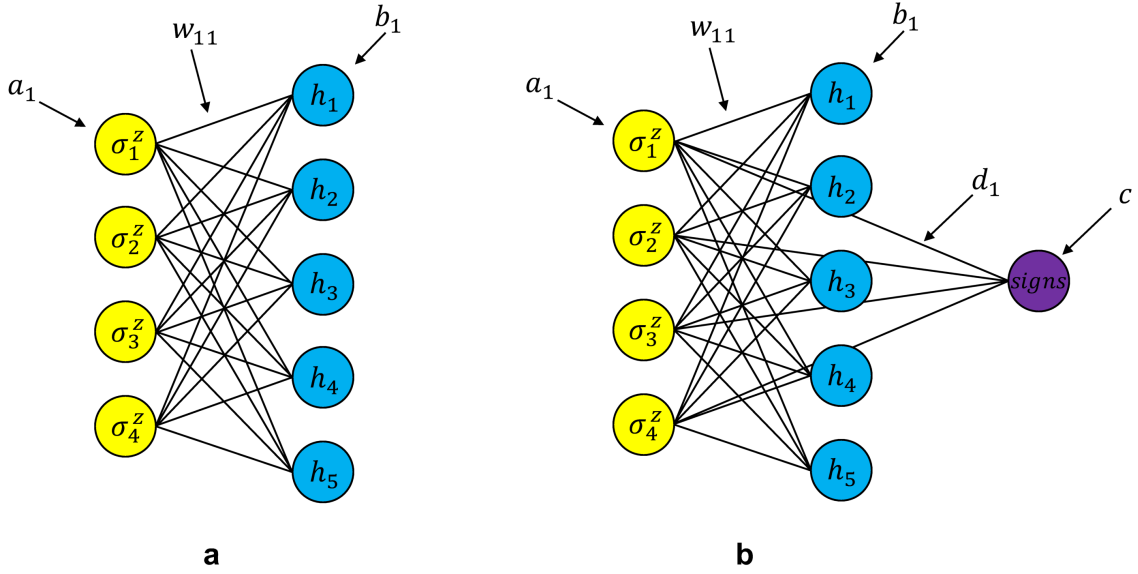}
\caption{Constructions of restricted Boltzmann machine. {\bf a}: the original restricted Boltzmann machine (RBM) structure with visible ${\sigma^z}$ and hidden $h$ layers. {\bf b}: Improved RBM structure with three layers, visible, hidden and sign. ${a_i,w_{ij}, b_i,d_i,c}$ are trainable weights describing the different connection between layers.}
\end{figure}

The $s$ uses a non-linear function $tanh$ to classify whether the sign should be positive or negative.  Because we have added another function for the coefficients, the distribution is not solely decided by RBM. We also need to add our sign function into the distribution.  Within this scheme, $c$ is a regulation and $d_i$ are weights for $\z{i}$. (see Figure 1). Our final objective function, now with $|\phi\rangle = \sum_{x}\phi(x)s(x)|x\rangle$, becomes:

\begin{equation}
\langle H\rangle =\frac{\sum_{x,x'}\overline{\phi(x)}\overline{s(x)}\langle x|H|x'\rangle \phi(x')s(x')}{\sum_x|\phi(x) s(x)|^2}
\end{equation}

After setting the objective function, the learning procedure is performed by sampling to get the distribution of $\phi(x)$ and calculating to get $s(x)$. We then proceed to calculate the joint distribution determined by $\phi(x)$ and $s(x)$. The gradients are determined by the joint distribution and we use gradient decent method to optimize $\langle H \rangle$ (see Supplementary
Note 1). Calculating the the joint distribution is efficient because $s(x)$ is only related to $x$.

{\bf Electronic Structure Hamiltonian Preparation.} The electronic structure is represented by $N$ single-particle orbitals which can be empty or occupied by a spinless electron\cite{lanyon2010towards}:

\begin{equation}
   \hat H= \sum_{i,j}h_{ij}a_i^\dagger a_j+\frac{1}{2}\sum_{i,j,k,l} h_{ijkl}a_i^\dagger a_j^\dagger a_ka_l
\end{equation}

where $h_{ij}$ and $h_{ijkl}$ are one and two-electron integrals. In this study we use the minimal basis (STO-3G) to calculate them. $a_j^{\dagger}$ and $a_j$ are creation and annihilation operators for the orbital $j$.

Equation (5) is then transformed to Pauli matrices representation, which is achieved by the Jordan-Wigner transformation\cite{fradkin1989jordan}. The final electronic structure Hamiltonian takes the general form with $\sigma_\alpha^i \in \left\{\sigma_x,\sigma_y,\sigma_z,I\right\}$ where $\sigma_x,\sigma_y,\sigma_z$ are Pauli matrices and $I$ is the identity matrix\cite{xia2017electronic}:

\begin{equation}
H=\sum\limits_{i,\alpha}h_\alpha^i\sigma_\alpha^i+\sum\limits_{i,j,\alpha,\beta}h_{\alpha\beta}^{ij}\sigma_\alpha^i\sigma_\beta^j+\sum\limits_{i,j,k,\alpha,\beta,\gamma}h_{\alpha\beta\gamma}^{ijk}\sigma_\alpha^i\sigma_\beta^j\sigma_\gamma^k+...
\end{equation}

{\bf Quantum algorithm to sample Gibbs distribution.} We propose a quantum algorithm to sample the distribution determined by RBM. The probability for each combination $y =\{\sigma^z,\ h\}$ can be written as:

\begin{equation}
P(y) = \frac{e^{\sum_{i}a_i\z{i}+\sum_{j}b_jh_j+\sum_{i,j}w_{ij}\z{i}h_j}}{\sum_{y'}e^{\sum_{i}a_{i}\Z{i}+\sum_{j}b_{j}h_{j}'+\sum_{i,j}w_{ij}\Z{i}h_{j}'}} 
\end{equation}

Instead of $P(y)$, we try to sample the distribution $Q(y)$ as:

\begin{equation}
Q(y) = \frac{e^{\frac{1}{k}(\sum_{i}a_i\z{i}+\sum_{j}b_jh_j+\sum_{i,j}w_{ij}\z{i}h_j)}}{\sum_{y'}e^{\frac{1}{k}(\sum_{i}a_{i}\Z{i}+\sum_{j}b_{j}h_{j}'+\sum_{i,j}w_{ij}\Z{i}h_{j}')}} 
\end{equation}

where $k$ is an adjustable constant with different values for each iteration and is chosen to increase the probability of successful sampling. In our simulation, it is chosen as $O(\sum_{i,j}|w_{ij}|)$.

We employed a quantum algorithm to sample the Gibbs distribution from the quantum computer. This algorithm is based on sequential applications of controlled-rotation operations, which tries to calculate a distribution $Q'(y) \geq Q(y)$ with an ancilla qubit showing whether the sampling for $Q(y)$ is successful\cite{wiebe2016quantum}.

This two-step algorithm uses one system register (with $n+m$ qubits in use) and one scratchpad register (with one qubit in use) as shown in Figure 2. 

All qubits are initialized as $|0\rangle$ at the beginning. The first step is to use $R_y$ gates to get a superposition of all combinations of $\{\sigma^z,h\}$ with $\theta_i = 2arcsin(\sqrt{\frac{e^{a_i/k}}{e^{a_i/k}+e^{-a_i/k}}})$ and $\gamma_j=2arcsin(\sqrt{\frac{e^{b_j/k}}{e^{b_j/k}+e^{-b_j/k}}})$:

\begin{equation*}
    \otimes_{i}R_y(\theta_i)|0_i\rangle\otimes_{j}R_y(\gamma_j)|0_j\rangle|0\rangle =
\textstyle\sum_{y} \sqrt{O(y)}|y\rangle|0\rangle
\end{equation*}

where $O(y) = \frac{e^{\sum_{i}a_i\z{i}/k+\sum_{j}b_jh_j/k}}{\sum_{y'}e^{\sum_{i}a_{i}\Z{i}/k+\sum_{j}b_{j}h_{j}'/k}}$ and $|\phi_y\rangle$ corresponds to the combination $|y\rangle =|\z{1}..\z{n}h_1...h_m\rangle$.

The second step is to calculate $e^{w_{ij}\z{i}h_j}$. We use controlled-rotation gates to achieve this. The idea of sequential controlled-rotation gates is to check whether the target qubit is in state $|0\rangle$ or state $|1\rangle$ and then rotate the corresponding angle (Figure 2). If qubits $\z{i}$ and $h_j$ are in $|00\rangle$ or $|11\rangle$, the ancilla qubit is rotated by $R_y(\theta_{ij,1})$ and otherwise by $R_y(\theta_{ij,2})$, with $\theta_{ij,1} = 2arcsin(\sqrt{\frac{e^{w_{ij}/k}}{e^{|w_{ij}|/k}}})$ and $\theta_{ij,2} = 2arcsin(\sqrt{\frac{e^{-w_{ij}/k}}{e^{|w_{ij}|/k}}})$. Each time after one $e^{w_{ij}\z{i}h_j}$ is calculated, we do a measurement on the ancilla qubit. If it is in $|1\rangle$ we continue with a new ancilla qubit initialized in $|0\rangle$ , otherwise we start over from the beginning (details in  Supplementary Note 2).

\begin{figure}[H]
\centering
\includegraphics[width=4in]{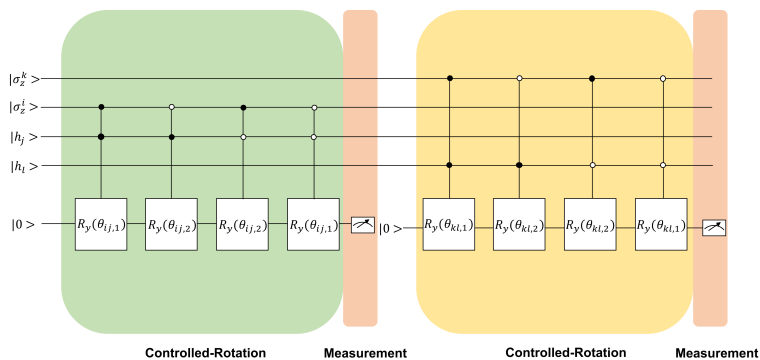}
\caption{The example circuit for the controlled-rotation gate approach with measurements.}
\end{figure}

After we finish all measurements the final states of the first $m+n$ qubits follow the distribution $Q(y)$.  We just measure the first $n+m$ qubits of the system register to obtain the probability distribution. After we get the distribution, we calculate all probabilities to the power of $k$ and normalize to get the Gibbs distribution.

\begin{figure}[H]
\centering
\includegraphics[width=4in]{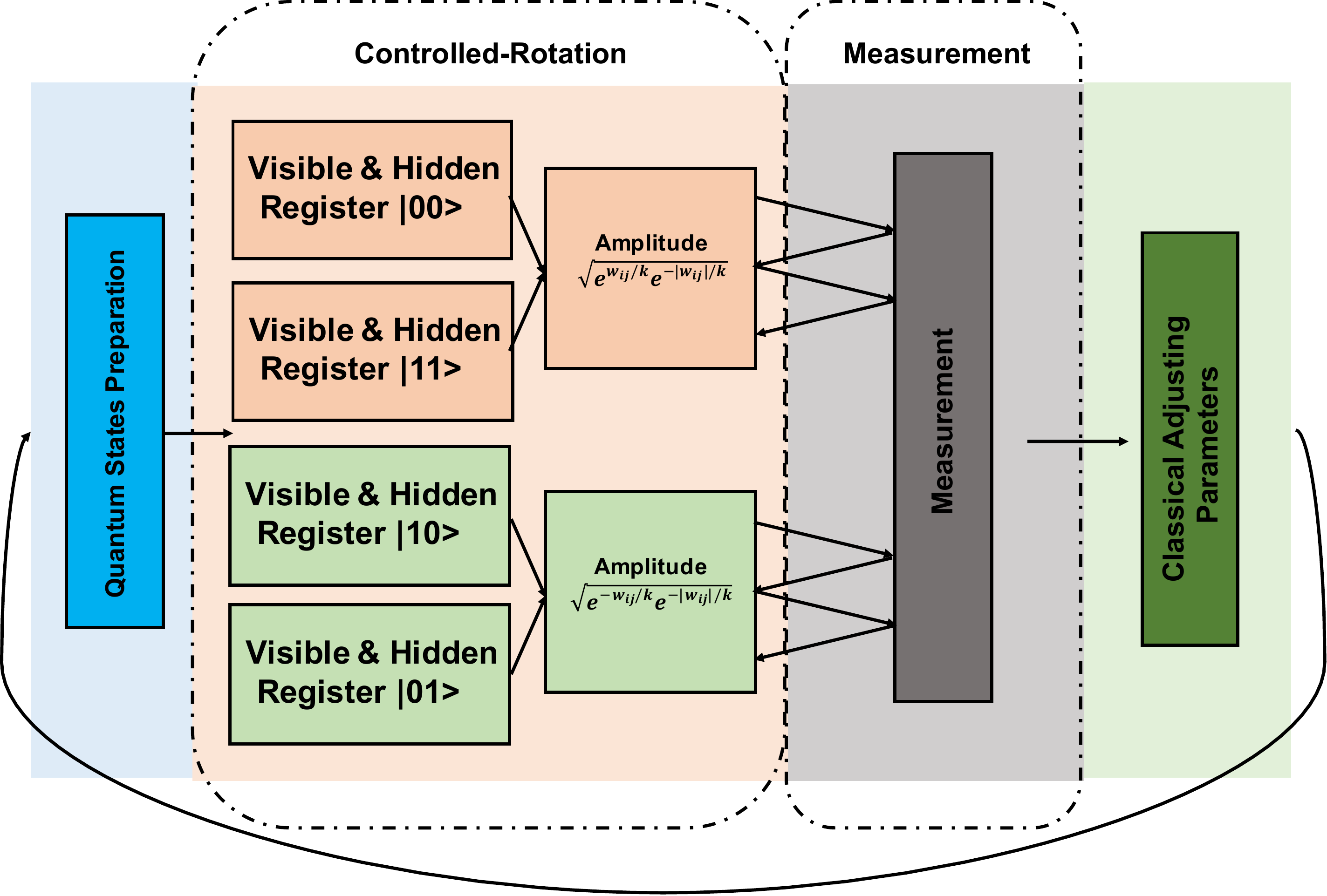}
\caption{The algorithmic flow chart of the quantum algorithm based on sequential controlled-rotations gates.}
\end{figure}

The complexity of gates comes to $O(mn)$ for one sampling and the qubits requirement comes to $O(mn)$. If considering the reuse of ancilla qubits, the qubits requirements reduce to $O(m+n)$ (see Supplementary Note 4). The probability of one successful sampling has a lower bound $e^{\frac{-1}{k}\sum_{i,j}2|w_{ij}|}$ and if $k$ is set to $O(\sum_{i,j}|w_{ij}|)$ it has constant lower bound (see Supplementary Note 3). If $N_s$ is the number of successful sampling to get the distribution, the complexity for one iteration should be $O(N_{s}mn)$ due to the constant lower bound of successful sampling as well as processing distribution taking $O(N_{s})$. In the meantime, the exact calculation for the distribution has complexity as $O(2^{m+n})$.  The only error comes from the error of sampling if not considering noise in the quantum computer.

{\bf Summary of numerical results.} We now present the results derived from our RBM for H$_2$, LiH and H$_2$O molecules. It can clearly be seen from Figure 4 that our three layer RBM yields very accurate results comparing to the disorganization of transformed Hamiltonian which is calculated by a finite minimal basis set, STO-3G. Points deviating from the ideal curve are likely due to local minima trapping during the optimization procedure. This can be avoided in the future by implementing optimization methods which include momentum or excitation, increasing the escape probability from any local features of the potential energy surface. 

Further discussion about our results should mention instances of transfer learning. Transfer learning is a unique facet of neural network machine learning algorithms describing an instance (engineered or otherwise) where the solution to a problem can inform or assist in the solution to another similar subsequent problem. Given a diatomic Hamiltonian at a specific intermolecular separation, the solution yielding the variational parameters --- which are the weighting coefficients of the basis functions --- are adequate first approximations to those parameters at a subsequent calculation where the intermolecular separation is a small perturbation to the previous value.

Except for the last point in the Figure 4 d, we use  1/40 of the iterations  for the last point in calculations initiated with transferred parameters from previous iterations of each points and still achieve a good result. We also see that the local minimum is avoided if the starting point achieve global minimum.

\begin{figure}[H]
\centering
\includegraphics[width=6in]{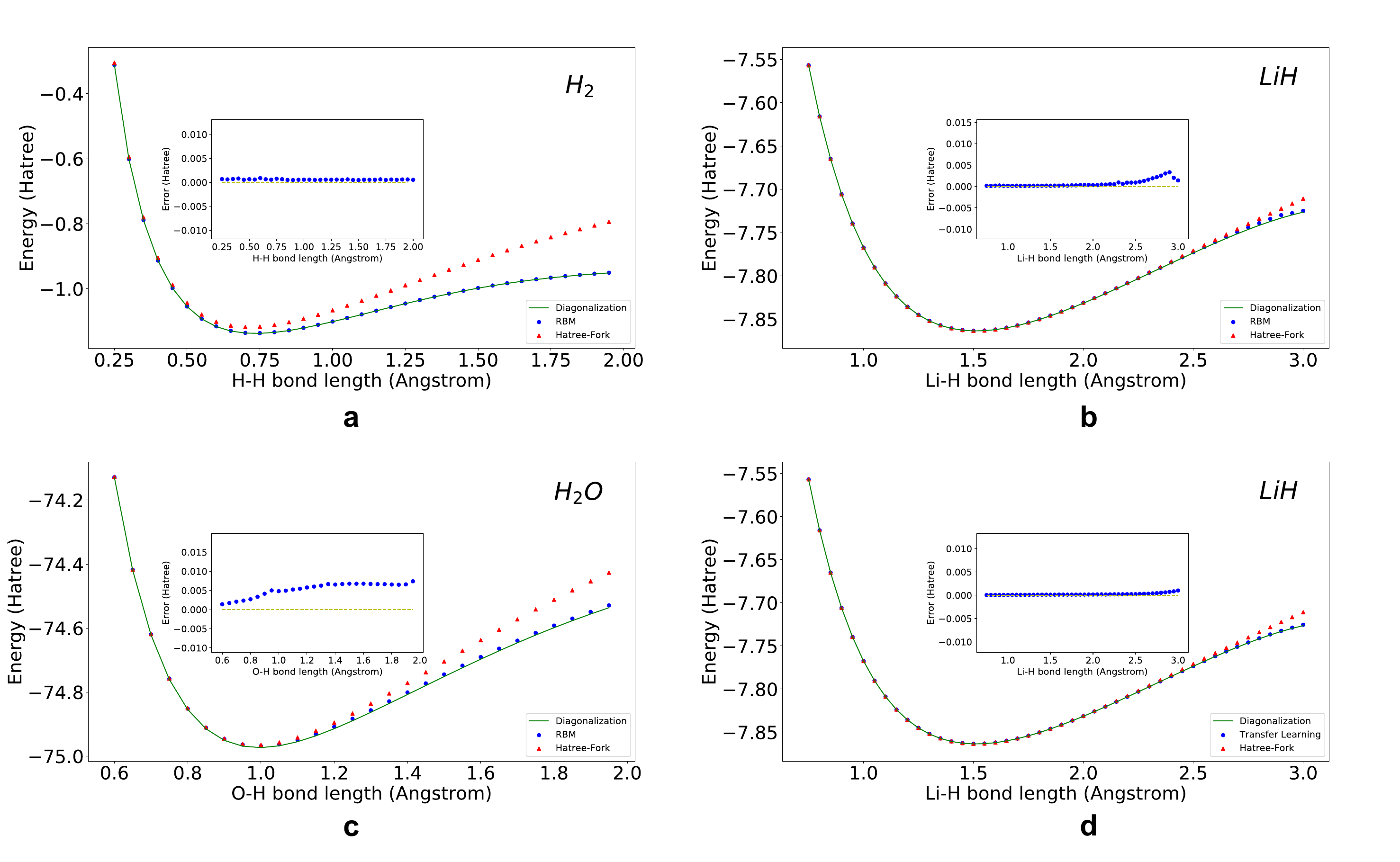}
\caption{Results of calculating ground state energy of H$_2$, LiH and H$_2$O. {\bf a}, {\bf b}, {\bf c} are the results of H$_2$ ($n=4, \ m=8$), LiH ($n=4, \ m=8$) and H$_2$O ($n=6, \ m=6$) calculated by our three layer RBM compared with exact diagonalized results of the transformed Hamiltonian. {\bf d} is the result of LiH ($n=4, \ m=8$) calculated by the Transfer Learning method. We use STO-3G as basis to compute the molecular integrals for the Hamiltonian. Bond length represents inter-atomic distance for the diatomic molecules and the distance O-H of the optimized equilibrium structure of the water molecule. The data points of RBM are minimum energies of all energies calculated during the whole optimization by sampling.}
\end{figure}

{\bf Discussion.} In conclusion, we present a combined quantum machine learning approach to perform electronic structure calculations. Here, we have a proof of concept and show results for small molecular systems. Screening molecules to accelerate the discovery of new materials for specific application is demanding since the chemical space is very large! For example, it was reported that the total number of possible
small organic molecules that populate the ‘chemical space’ 
exceed $10^{60}$\cite{dobson2004chemical,blum2009970}. Such an enormous size makes a thorough exploration of chemical
space using the traditional electronic structure methods impossible. Moreover, in a recent perspective\cite{p}in Nature Reviews Materials the potential of machine learning algorithms to accelerate the discovery of materials was pointed out.  Machine learning algorithms have been used for material screening.
For example, out of the GDB-17 data base, consisting of about 166 billion molecular graphs, one can make organic and drug-like molecules with up to 17 atoms and 134 thousand smallest molecules with up to 9 heavy atoms were calculated using hybrid density functional (B3LYP/6-31G(2df,p). Machine learning algorithms trained on these data, were found to predict molecular properties of subsets of these molecules \cite{tabor2018accelerating,von2018quantum,ramakrishnan2014quantum}.  

In the current simulation, H$_2$ requires 13 qubits with the number of visible units $n=4$, the number of hidden units $m=8$ and additional 1 reusing ancilla qubits . LiH requires 13 qubits with the number of visible units $n=4$, the number of hidden units $m=8$ and additional 1 reusing ancilla qubits. H$_2$O requires 13 qubits with the number of visible units $n=6$, the number of hidden units $m=6$ and additional 1 reusing ancilla qubits. The order of scaling of qubits for the system should be $O(m+n)$ with reusing ancilla qubits. The number of visible units $n$ is equal to the number of spin orbitals. The choice of the number of hidden units $m$ is normally integer times of $n$ which gives us a scaling of $O(n)$ with reusing ancilla qubits . Thus, the scaling of the qubits increases polynomially with the number of spin orbitals. Also, the complexity of gates $O(n^2)$ scales polynomially with the number of spin orbitals while the scaling of classical Machine Learning approaches calculating exact Gibbs distribution is exponential. With the rapid development of larger-scale quantum computers and the possible training of some machine units with the simple dimensional scaling results for electronic structure, quantum machine learning techniques are set to become powerful tools to perform electronic structure calculations and assist in designing new materials for specific applications.

\section*{Methods}

{\bf Preparation of the Hamiltonian of H$_2$, LiH and H$_2$O.} We treat H$_2$ molecule with 2-electrons in a minimal basis STO-3G and use the Jordan-Wigner transformation\cite{fradkin1989jordan}. The final Hamiltonian is of 4 qubits.  We treat LiH molecule with 4-electrons in a minimal basis STO-3G and use the Jordan-Wigner transformation\cite{fradkin1989jordan}. We assumed the first two lowest orbitals are occupied by electrons and the the final Hamiltonian is of $4$ qubits.  We treat H$_2$O molecule with 10-electrons in a minimal basis STO-3G, we use Jordan-Wigner transformation\cite{fradkin1989jordan}. We assume the first four lowest energy orbitals are occupied by electrons and first two highest energy orbitals are not occupied all time. We also use the spin symmetry in \cite{kandala2017hardware,bravyi2017tapering} to reduce another two qubits. With the reduction of  the number of qubits, finally we have $6$ qubits Hamiltonian \cite{xia2017electronic, bian2018comparison}. All calculations of integrals in second quantization and transformations of electronic structure are done by OpenFermion\cite{mcclean2017openfermion} and Psi4\cite{parrish2017psi4}.

{\bf Gradient estimation.}
The two functions $\phi(x)$ and $s(x)$ are both real function. Thus, the gradient for parameter $p_k$ can be estimated as  $2(\langle E_{loc}D_{p_k}\rangle -\langle E_{loc}\rangle\langle D_{p_k}\rangle)$ where $E_{loc}(x)=\frac{\langle x|H|\phi\rangle}{\phi(x)s(x)}$ is so called local energy, $D_{p_k}(x) = \frac{\partial_{p_k}(\phi(x)s(x))}{\phi(x)s(x)}$. $\langle...\rangle$ represents the expectation value of joint distribution determined by $\phi(x)$ and $s(x)$ (details in  Supplementary Note 1).

{\bf Implementation Details.}
In our simulation we choose small constant learning rate 0.01 to avoid trapping in local minimum. All parameter are initialized as a random number between $(-0.02, 0.02)$. The range of initial random parameter is to avoid gradient vanishing of $tanh$. For each calculation we just need 1 reusing ancilla qubit all the time.  Thus, in the simulation, the number of required qubits is $m+n+1$.  All calculations do not consider the noise and system error (details in  Supplementary Note 5).

{\bf Data availability.} The data and codes that support the findings of this study are available
from the corresponding author upon reasonable request.

\bibliographystyle{unsrt}
\bibliography{main.bib}

\section*{Acknowledgement}
We would thank to Dr. Ross Hoehn, Dr. Zixuan Hu and Teng Bian for critical reading and useful discussions. S.K and R.X are grateful for the support from Integrated Data Science Initiative Grants, Purdue University.

\section*{Author Contribution}
S.K designed the research. R.X performed the calculations. Both discussed the results and wrote the paper. 

\section*{Additional Information}
{\bf Supplementary Information.} Supplementary Materials are available.

{\bf Competing Interests.} The authors declare no competing interest.

\newpage

\section*{Supplementary Note 1}
\subsection*{Derivation of the gradient}
For an electronic structure Hamiltonian prepared by second quantization and Jordan-Wigner transformation\cite{fradkin1989jordan}, $H$, and a trial wave function, $|\phi\rangle= \sum_{x}\phi(x)s(x)|x\rangle$, the expectation value can be written as\cite{carleo2018neural}:

\begin{equation}
\langle H\rangle=\frac{\langle\phi|H|\phi\rangle}{\langle\phi|\phi\rangle}=\frac{\sum_{x,x'}\overline{\phi(x)}\overline{s(x)}\langle x|H|x'\rangle \phi(x')s(x')}{\sum_x{|\phi(x)s(x)|^2}}
\end{equation}

$x$ is a combination of  $\{\z{1},\z{2}...\z{n}\}$ and $|x\rangle=|\z{1}\z{2}...\z{n}\rangle$.

If we set $\Phi(x) = \phi(x)s(x)$, because $\phi(x)$ and $s(x)$ are all real value functions, then the gradient can be calculated as\cite{carleo2018neural,carleo2017solving}:

\begin{equation}
    \begin{aligned}
        \partial_{p_k}\langle H\rangle &=\frac{\sum_{x}(\partial_{p_k}\Phi(x))\langle x|H|\phi\rangle+\sum_{x}\langle \phi|H|x\rangle(\partial_{p_k}\Phi(x))}{\sum_x|\Phi(x)|^2}\\&-\frac{\sum_{x}\Phi(x)\langle x|H|\phi\rangle}{\sum_x|\Phi(x)|^2}\frac{\sum_x((\partial_{p_k}\Phi(x))\Phi(x)+\Phi(x)\partial_{p_k}\Phi(x))}{\sum_x|\Phi(x)|^2}\\
    \end{aligned}
\end{equation}

If we set $E_{loc}(x)=\frac{\langle x|H|\phi\rangle}{\Phi(x)}$ and $ D_{p_k}(x)=\frac{\partial_{p_k}\Phi(x)}{\Phi(x)}$, the gradient can be written as\cite{carleo2018neural}:

\begin{equation}
\begin{aligned}
     \partial_{p_k}\langle H\rangle &=\frac{\sum_{x}D_{p_k}(x)E_{loc}(x)|\Phi(x)|^2+\sum_{x}E_{loc}(x)D_{p_k}(x)|\Phi(x)|^2}{\sum_x|\Phi(x)|^2}\\&-\frac{\sum_{x}|\Phi(x)|^2E_{loc}(x)}{\sum_x|\Phi(x)|^2}\frac{\sum_x(D_{p_k}(x)+D_{p_k}(x))|\Phi(x)|^2}{\sum_x|\Phi(x)|^2}\\&=2\langle E_{loc}D_{p_k}\rangle - 2\langle E_{loc}\rangle\langle D_{p_k}\rangle
\end{aligned}
\end{equation}

where $\langle...\rangle$ represent the expectation value of distribution determined by $\Phi(x)$. $\langle x|H|\phi\rangle = \langle \phi| H|x\rangle$ for that  $H$ is a real symmetric matrix due to Jordan-Wigner transformation.

$p_k$ is the parameters $a_i, b_j, w_{ij}, d_i, c$ for $k_{th}$ iterations. Thus we have\cite{carleo2018neural}:
\begin{equation}
\begin{aligned}
&D_{a_i}(x)=\frac{1}{2}\z{i} -\frac{1}{2}\langle\z{i}\rangle_{RBM},\\
&D_{b_j}(x)=\frac{1}{2}tanh(\theta_j)-\frac{1}{2}\langle h_j\rangle_{RBM},\\
&D_{w_{ij}}(x)=\frac{1}{2}tanh(\theta_j)\z{i}-\frac{1}{2}\langle\z{i}h_j\rangle_{RBM},\\
&D_{c}(x)=1/s(x)-s(x),\\
&D_{d_i}(x)=\z{i}(1/s(x)-s(x)),
\end{aligned}
\end{equation}

where $\theta_j=\sum_iw_{ij}\sigma^z_i+b_j$. $\langle...\rangle_{RBM}$ represents the distribution determined solely by RBM. We do not need to calculate the second term of $D_{a_i}$, $D_{b_i}$ and $D_{w_{ij}}$ for that they will be cancelled when calculating the gradient $ \partial_{p_k}\langle H\rangle$. We use the gradient decent method to optimize our RBM, yielding the global minimum corresponding to the ground energy.

\begin{equation}
p_{k+1}=p_{k}-\alpha_k\partial_{p_k}\langle H\rangle
\end{equation}

Where $\alpha_k$ is the learning rate for $k_{th}$ iteration, controlling the convergence rate. We can continue iterating until we reach the maximum number of iterations. The gradient is estimated by the distribution calculated by sampling.

\section*{Supplementary Note 2}

\subsection*{Sequential applications of controlled-rotation algorithm}

The probability for each combination $y =\{\sigma^z,\ h\}$ can be written as:

\begin{equation}
P(y) = \frac{e^{\sum_{i}a_i\z{i}+\sum_{j}b_jh_j+\sum_{i,j}w_{ij}\z{i}h_j}}{\sum_{y'}e^{\sum_{i}a_{i}\Z{i}+\sum_{j}b_{j}h_{j}'+\sum_{i,j}w_{ij}\Z{i}h_{j}'}}
\end{equation}

However, we do not directly calculate the $P(y)$  but we do some modification on $P(y)$ to increase the successful probability of our algorithm. We calculate $Q(y) =\frac{e^{\sum_{i}a_i\z{i}/k+\sum_{j}b_jh_j/k+\sum_{i,j}w_{ij}\z{i}h_j/k}}{\sum_{y'}e^{\sum_{i}a_{i}\Z{i}/k+\sum_{j}b_{j}h_{j}'/k+\sum_{i,j}w_{ij}\Z{i}h_{j}'/k}}$ where $k$ is a large number to increase the successful probability of our measurements.

First we use $R_y$ gate to achieve a superposition of all possible $\sigma^z$ and $h$. The system qubits and the ancilla qubit are initialized at state $|0\rangle$.

\begin{equation}
    \otimes_{i}R_y(2arcsin(\sqrt{\frac{e^{a_i/k}}{e^{a_i/k}+e^{-a_i/k}}))}|0_i\rangle\otimes_{j}R_y(2arcsin(\sqrt{\frac{e^{b_j/k}}{e^{b_j/k}+e^{-b_j/k}}}))|0_j\rangle|0\rangle=\sum_{y}\sqrt{O(y)}|y\rangle|0\rangle
\end{equation}

where $O(y) = \frac{e^{\sum_{i}a_i\z{i}/k+\sum_{j}b_jh_j/k}}{\sum_{y'}e^{\sum_{i}a_{i}\Z{i}/k+\sum_{j}b_{j}h_{j}'/k}}$ and $|y\rangle =|\z{1}..\z{n}h_1...h_m\rangle$.

The next step is to calculate each term of $e^{\sum_{i,j}w_{ij}\z{i}h_j}$, which is achieved by controlled rotations gates. The idea is, for each time controlled rotation, we calculate two angles $\theta_{ij,1} = 2arcsin(\sqrt{e^{w_{ij}/k}e^{-|w_{ij}|/k}})$ and $\theta_{ij,2} = 2arcsin(\sqrt{e^{-w_{ij}/k}e^{-|w_{ij}|/k}})$. We use controlled-rotation $CR_y(\theta_{ij,1})$ and $CR_y(\theta_{ij,2})$ which are controlled by combination of $\z{i}$, $h_j$ as working qubits  to do rotation on the ancilla qubit. The controlled rotation is to check the working qubits and then do the corresponding rotation $\theta_{ij,1}$ or $\theta_{ij,2}$.

All controlled rotation gates can be expressed as below:

\begin{equation}
\begin{aligned}
&CR_{w_{ij},1}=C_{\z{i},h_j}\otimes R_y(2arcsin(\sqrt{e^{w_{ij}/k}e^{-|w_{ij}|/k}}))+(D_{\z{i},h_j}+E_{\z{i},h_j}+F_{\z{i},h_j})\otimes I \\
&CR_{w_{ij},2}=D_{\z{i},h_j}\otimes R_y(2arcsin(\sqrt{e^{-w_{ij}/k}e^{-|w_{ij}|/k}}))+(C_{\z{i},h_j}+E_{\z{i},h_j}+F_{\z{i},h_j})\otimes I\\
&CR_{w_{ij},3}=E_{\z{i},h_j}\otimes R_y(2arcsin(\sqrt{e^{-w_{ij}/k}e^{-|w_{ij}|/k}}))+(C_{\z{i},h_j}+D_{\z{i},h_j}+F_{\z{i},h_j})\otimes I\\
&CR_{w_{ij},4}=F_{\z{i},h_j}\otimes R_y(2arcsin(\sqrt{e^{w_{ij}/k}e^{-|w_{ij}|/k}}))+(C_{\z{i},h_j}+D_{\z{i},h_j}+E_{\z{i},h_j})\otimes I\\
\end{aligned}
\end{equation}

where $C_{\z{i},h_j} = B_{\z{i}}\otimes B_{h_j}$, $D_{\z{i},h_j} = A_{\z{i}}\otimes B_{h_j}$, $E_{\z{i},h_j} = B_{\z{i}}\otimes A_{h_j}$, $F_{\z{i},h_j} = A_{\z{i}}\otimes A_{h_j}$ and  

\[
A=
\begin{bmatrix}
1 & 0\\
0 & 0\\
\end{bmatrix} B=
\begin{bmatrix}
0 & 0\\
0 & 1\\
\end{bmatrix}
\] 

Between the calculation of two $w_{ij}$, we need to do a measurement on the ancilla qubit to make sure the state of system qubits collapse to the wanted state. Measuring ancilla qubit in $|1\rangle$ means the state of system qubits collapse to the wanted state as we initialize the ancilla qubit in $|0\rangle$. 

We then do the measurement, if and only if the ancilla qubit is in $|1\rangle$ we continue with a new ancilla qubit initialized in $|0\rangle$, otherwise we start from beginning. The probability of success is very large since we choose $k$ as a large number.

After we finish all measurements, the distribution should be $Q(y)$. We just measure the first $n+m$ qubits of the system register to obtain the probability distribution. After we get the distribution, we calculate all probabilities to the power of $k$ and normalize to get the Gibbs distribution.

\section*{Supplementary Note 3}
\subsection*{Lower bound of successful sampling}
The successful probability $P$ can be written as:

\begin{equation}
\begin{aligned}
     P = \frac{\sum_{\sigma^z,h}e^{\frac{1}{k}(\sum_{i}a_i\z{i}+\sum_{j}b_jh_j+\sum_{i,j}w_{ij}\z{i}h_j)}}{\sum_{\sigma^z,h}e^{\frac{1}{k}(\sum_{i}a_i\z{i}+\sum_{j}b_jh_j)}e^{\frac{1}{k}(\sum_{i,j}|w_{ij}|)}}\geq \frac{e^{\frac{-1}{k}(\sum_{i,j}|w_{ij}|)}}{e^{\frac{1}{k}(\sum_{i,j}|w_{ij}|)}}=\frac{1}{e^{\frac{1}{k}(\sum_{i,j}2|w_{ij}|)}}
\end{aligned}
\end{equation}

If we choose $k = O(\sum_{i,j}|w_{ij}|)$, we have $P\geq\frac{1}{e^{O(1)}}$ which means the lower bound of probability of successful sampling is a constant. In the simulation, we choose $k=max(\frac{1}{2}\sum_{i,j}|w_{ij}|,1)$ because larger $k$  introduces larger sampling errors. This particular choice of $k$ gives us lower bound of success as $e^{-4}$. But in numerical simulation, the probability is much larger than $e^{-4}$, see {\bf Supplementary Note 5}.

\section*{Supplementary Note 4}
\subsection*{Complexity and error of the algorithm}
For a $C^2(U)$ conditioned by $|11\rangle$, it can be decomposed as the below\cite{nielsen2002quantum}:

\begin{figure}[H]
    \centering
    \includegraphics[width = 3in]{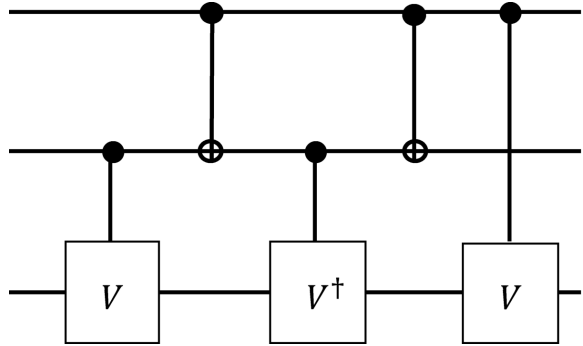}
    \caption*{{\bf Supplementary Figure 1}: The decomposition of the $C^2(U)$ gate.}
\end{figure}

where $V^2 = U$. In our algorithm, $U=R_y(\theta)$, thus we can choose $V=R_y(\theta/2)$ to achieve the decomposition. $C^2(U)$ conditioned  by $|00\rangle$, $|10\rangle$ or $|01\rangle$ can be achieved by adding $X$ gates on controlling qubits. For each $w_{ij}$, we have 4 $C^2(U)$ which means the gates complexity scales to $O(mn)$ and the number of qubits for our algorithms scales to $O(mn)$, which can be reduced to $O(m+n)$ if considering qubit reuse. Because the lower bound of probability of successful sampling is constant, if the number of successful sampling is $N_s$, the complexity for each iteration is $O(N_smn)$. The only error comes from the error of sampling if not considering noise in the quantum computer.

\section*{Supplementary Note 5}
\subsection*{Implementation details for H$_2$, LiH and H$_2$O}

Here we present the probabilities of successful sampling when calculating H$_2$, LiH and H$_2$O for bond length equals to 1.75 Angstrom.

\begin{figure}[H]
\begin{minipage}[t]{0.5\linewidth}
\centering
\includegraphics[width=3in, height = 2in]{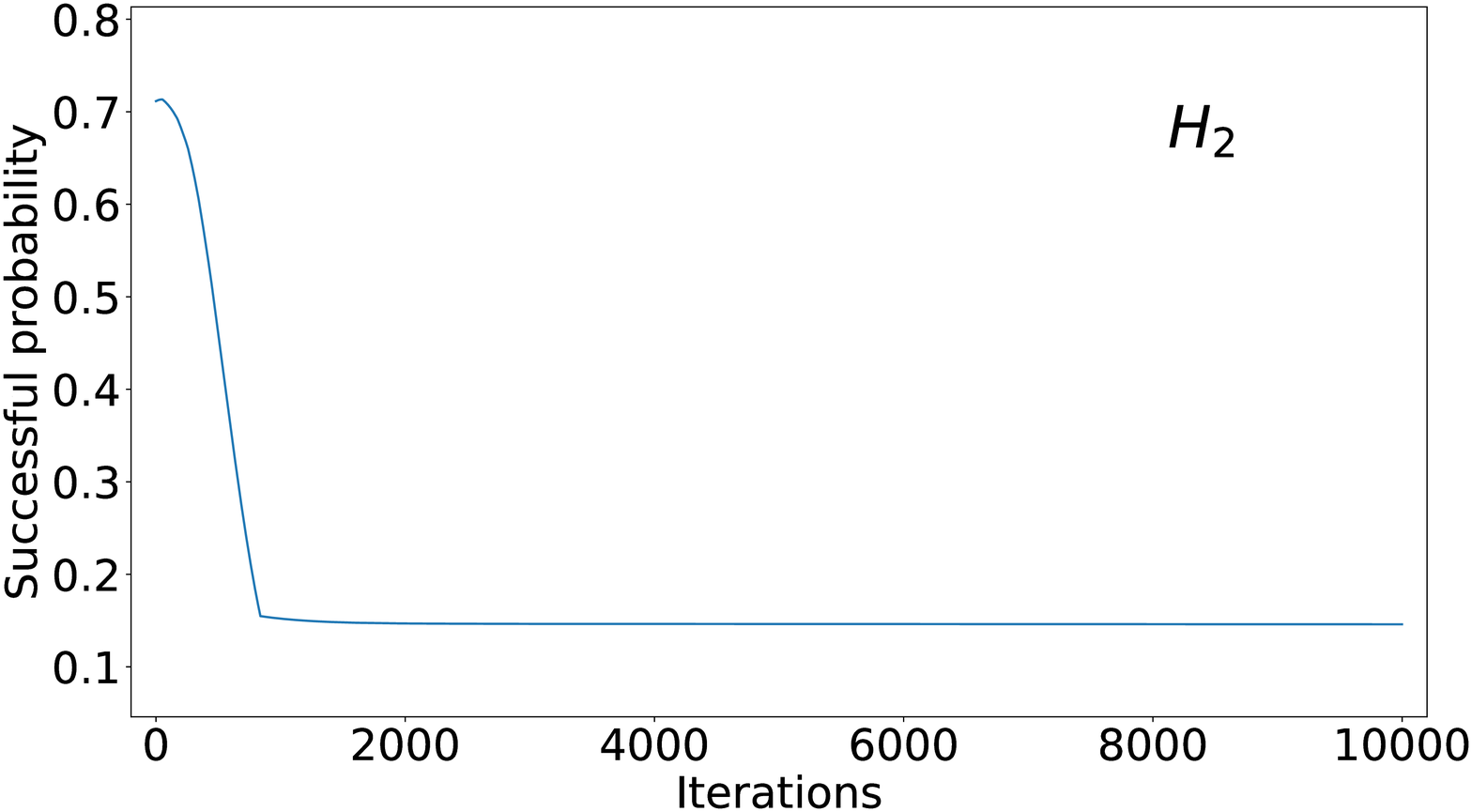}
\caption*{\bf{a}}
\end{minipage}%
\begin{minipage}[t]{0.5\linewidth}
\centering
\includegraphics[width=3in, height = 2in]{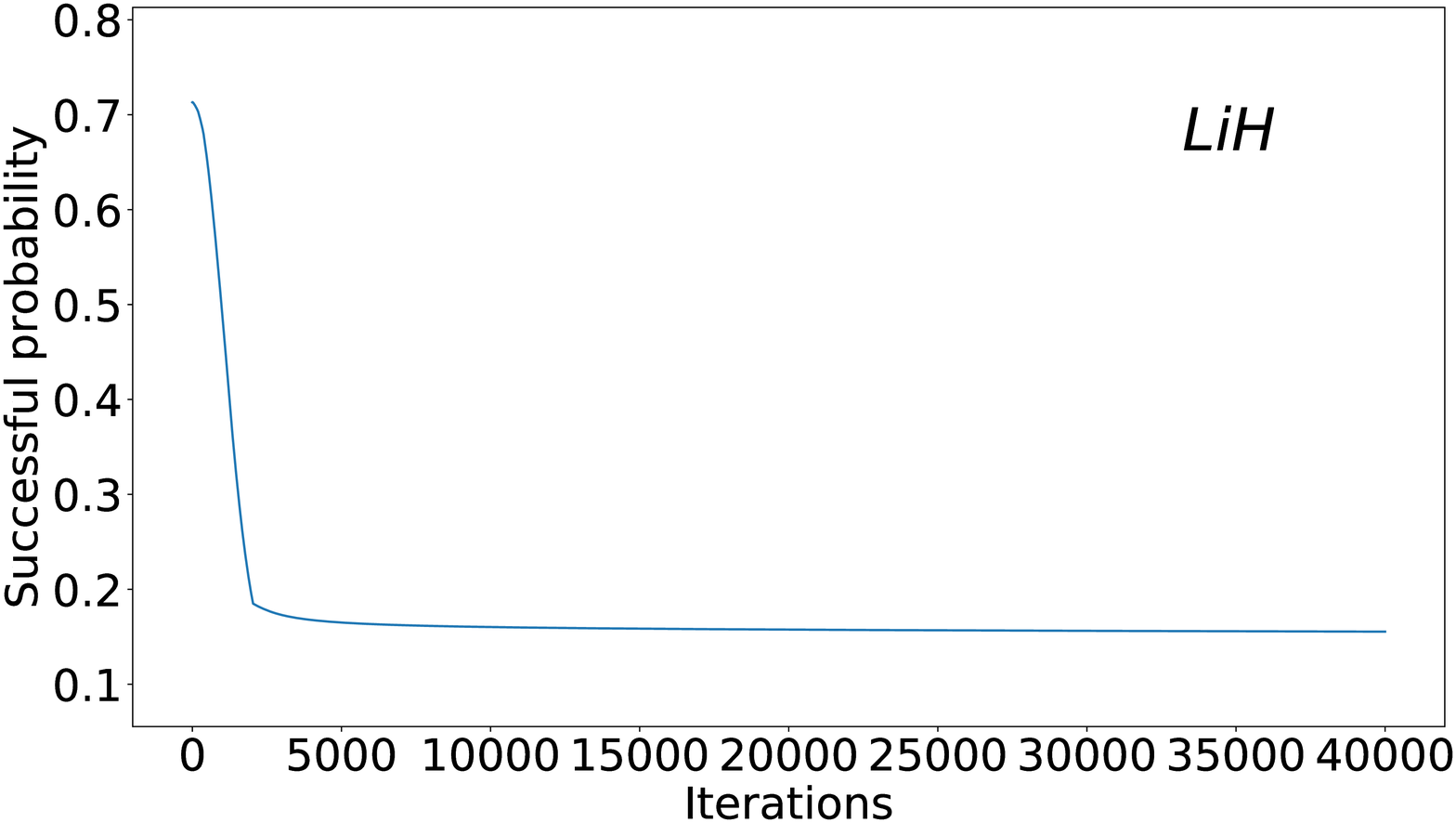}
\caption*{\bf{b}}
\end{minipage}
\begin{minipage}[t]{0.5\linewidth}
\centering
\includegraphics[width=3in, height = 2in]{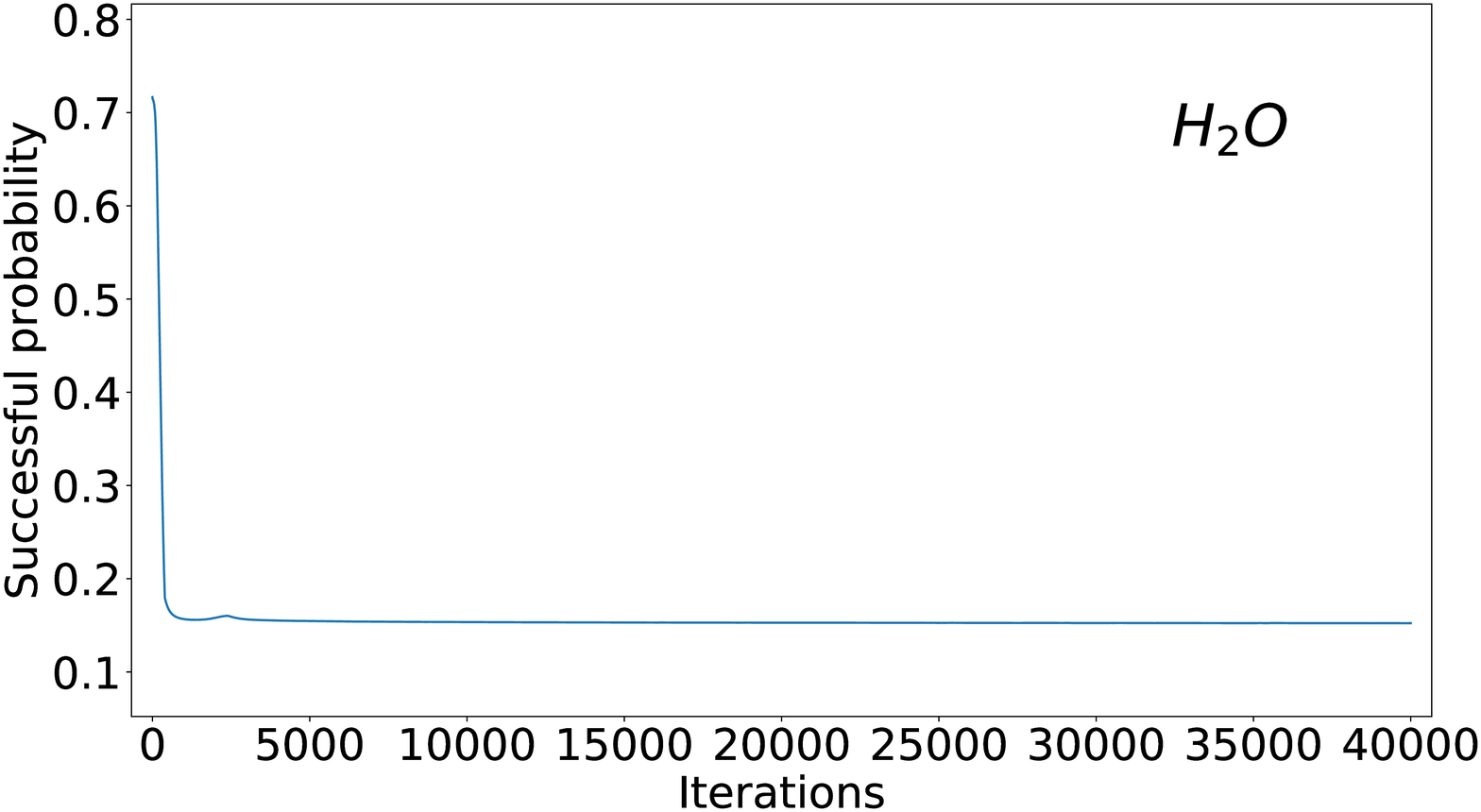}
\caption*{\bf{c}}
\end{minipage}%
\caption*{{\bf Supplementary Figure 2}: The probability of successful sampling during the optimization. {\bf a} Optimization procedure for H$_2$. {\bf b} Optimization procedure for LiH. {\bf c} Optimization procedure for H$_2$O. }
\end{figure}

Here we present the the changes of energy during optimization when calculating H$_2$, LiH and H$_2$O for bond length equals to 1.75 Angstrom..

\begin{figure}[H]
\begin{minipage}[t]{0.5\linewidth}
\centering
\includegraphics[width=3in, height = 2in]{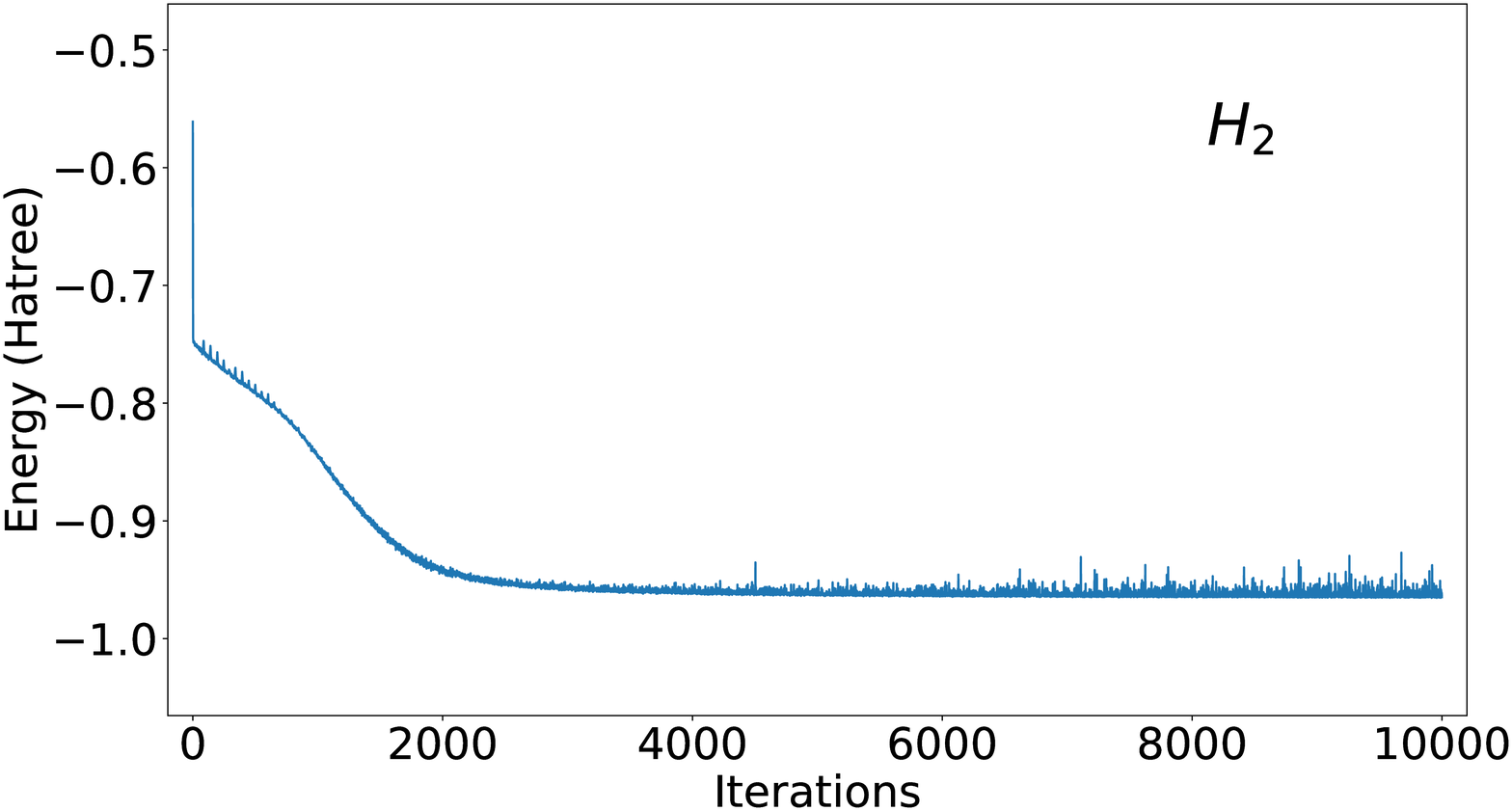}
\caption*{\bf{a}}
\end{minipage}%
\begin{minipage}[t]{0.5\linewidth}
\centering
\includegraphics[width=3in, height = 2in]{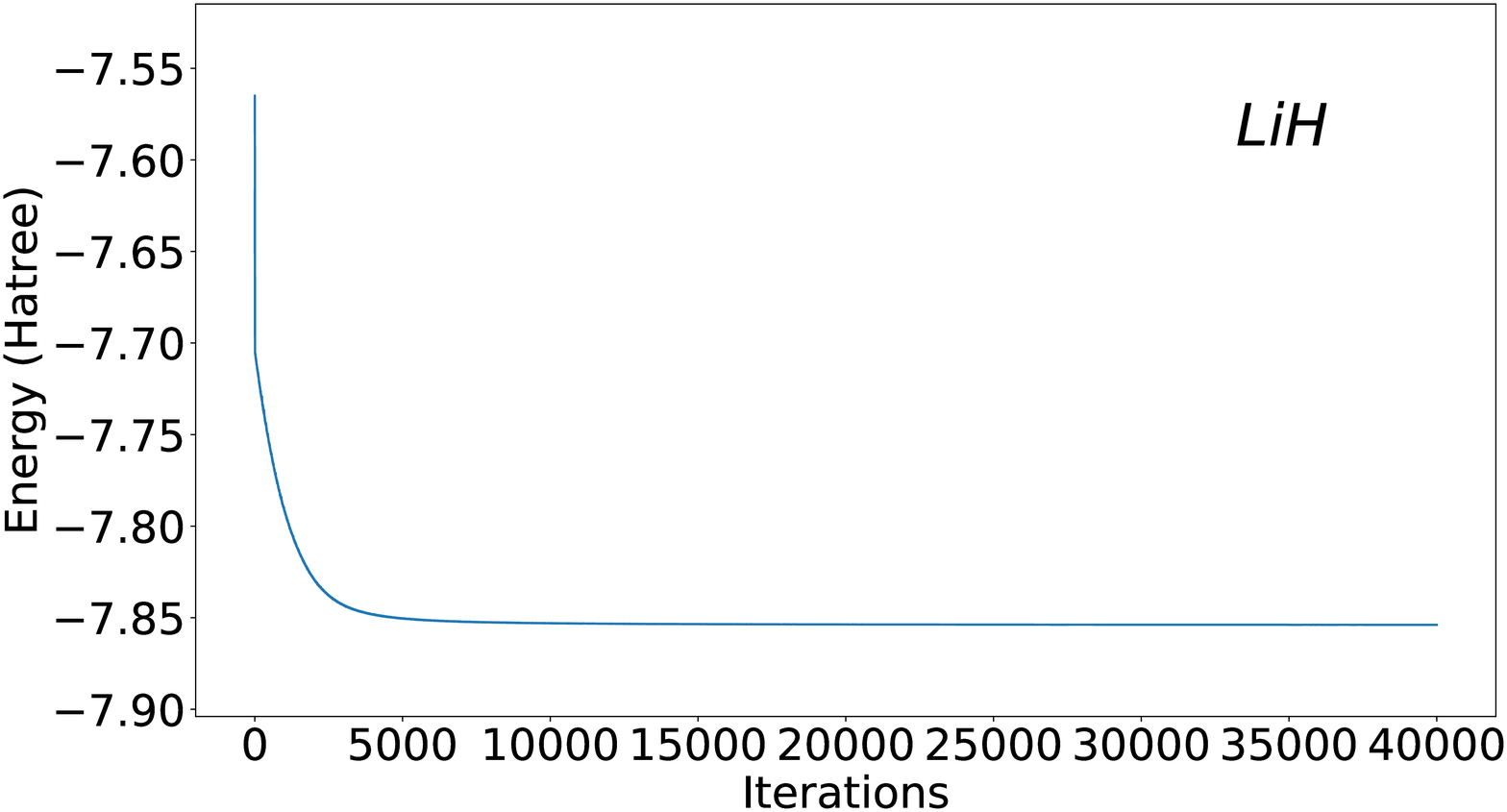}
\caption*{\bf{b}}
\end{minipage}
\begin{minipage}[t]{0.5\linewidth}
\centering
\includegraphics[width=3in, height = 2in]{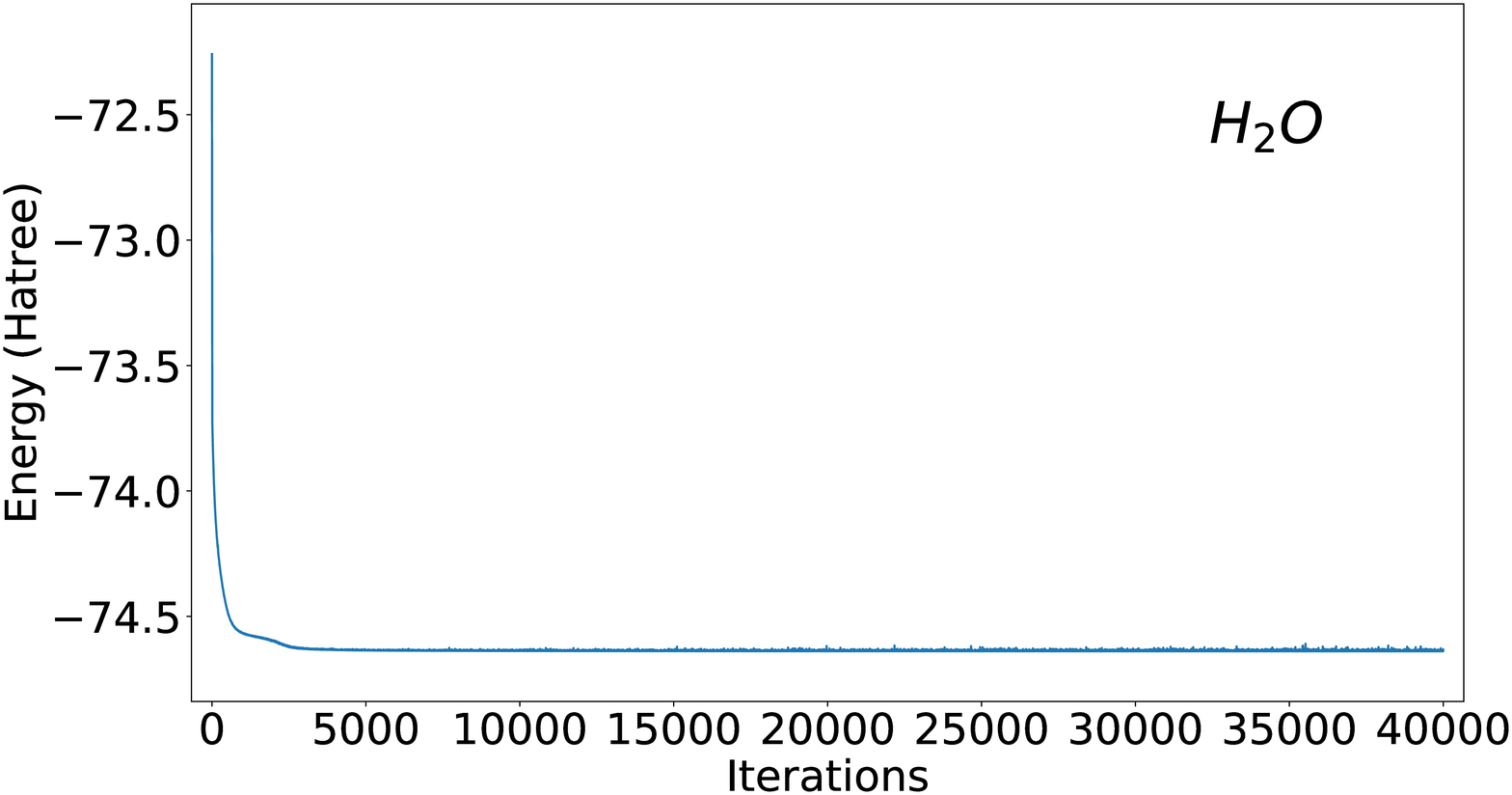}
\caption*{\bf{c}}
\end{minipage}%
\caption*{{\bf Supplementary Figure 3}: The energies calculated by RBM during the optimization. {\bf a} Optimization procedure for H$_2$. {\bf b} Optimization procedure for LiH. {\bf c} Optimization procedure for H$_2$O. }
\end{figure}

The distribution we want to sampling for the quantum algorithm is:

\begin{equation}
    Q(y) =\frac{e^{\sum_{i}a_i\z{i}/k+\sum_{j}b_jh_j/k+\sum_{i,j}w_{ij}\z{i}h_j/k}}{\sum_{y'}e^{\sum_{i}a_{i}\Z{i}/k+\sum_{j}b_{j}h_{j}'/k+\sum_{i,j}w_{ij}\Z{i}h_{j}'/k}}
\end{equation}

In our controlled-rotation algorithm, we use a $k$ as regulation to increase the probability of success as the proof in the {\bf Supplementary Note 3}, the lower bound of probability of success would become :

\begin{equation}
    \frac{1}{e^{\frac{1}{k}(\sum_{i,j}2|w_{ij}|)}}
\end{equation}

Thus, if no regulation ($k=1$), the probability of success would become $\frac{1}{e^{\sum_{i,j}2|w_{ij}|}}$ which means we need exponential number of measurements to get enough successful sampling, making no speedup in quantum algorithm.

If we add a regulation of $k$, in simulation we use $k=\frac{1}{2}\sum_{i,j}|w_{ij}|$, the probability becomes $\frac{1}{e^4}$ which needs constant number of measurements to get enough successful sampling.

After we get the distribution, we need to calculate all distribution to the power of $k$ and normalize to get the wanted distribution. $k$ is a large number, which is around $5$ at final for H$_2$, LiH and H$_2$O in our simulation. To decrease the errors in calculating power of $k$, we have to increase the number of sampling for our quantum algorithm when $k$ is large, which requires large number of sampling and may not be efficient. In the {\bf Supplementary Figure 3}, we can see that at the final procedure of optimization, the fluctuation is very large due to large $k$, which can be decreased by increasing the number of sampling. Because we investigated small molecule system H$_2$, LiH and H$_2$O, $k$ is not very large and the quantum algorithm is efficient. For large $k$, our quantum algorithm may require large sampling.

\end{document}